\documentclass[aip,pof,reprint]{revtex4-1}
\usepackage{graphicx,color,amsmath,amssymb}

\newcommand{\pd}{\partial}
\newcommand {\pdd}[2]{\frac{\partial #1}{\partial #2}}
\newcommand{\bv}[1] {\boldsymbol{{#1}}}   
\newcommand{\hv}[1]{\boldsymbol{\hat{{#1}}}} 
\newcommand{\sms}[1]{{_{_{#1}}}}  
\newcommand {\vel}{\bv{v}}
\newcommand{\xh}{\hv{x}}
\newcommand{\yh}{\hv{y}}
\def\grad{\bv{\nabla}}
\def\ubar{\overline{u}}

\def\HF{\mathcal{H}}
\def\biblio{OptimumNu}
\newcommand{\alv}{\alpha\sms{V}}
\newcommand{\beq}{\begin{equation}}
\newcommand{\eeq}{\end{equation}}

\begin{document}

\title{Heat Transport by Coherent Rayleigh-B\'enard Convection}
\author{Fabian Waleffe}
\affiliation{Department of Mathematics, University of Wisconsin, Madison, WI 53706}
\affiliation{Department of Engineering Physics, University of Wisconsin, Madison, WI 53706}
\email{waleffe@math.wisc.edu}
\author{Anakewit Boonkasame}
\affiliation{Department of Mathematics, University of Wisconsin, Madison, WI 53706}
\email{boonkasa@math.wisc.edu}
\author{Leslie M.\  Smith}
\affiliation{Department of Mathematics, University of Wisconsin, Madison, WI 53706}
\affiliation{Department of Engineering Physics, University of Wisconsin, Madison, WI 53706}
\email{lsmith@math.wisc.edu}

\date{\today}

\begin{abstract}

Steady but generally unstable solutions of the 2D Boussinesq equations are obtained for no-slip boundary conditions and Prandtl number 7.
The primary solution that bifurcates from the conduction state at Rayleigh number $Ra \approx 1708$ has been 
calculated up to  $Ra\approx 5. 10^6$ and its Nusselt number is  $Nu \sim 0.143\, Ra^{0.28}$ with a 
delicate  spiral structure in the temperature field. Another solution that maximizes $Nu$ over the horizontal wavenumber has been calculated up to $Ra=10^9$ and scales as  $Nu \sim 0.115\, Ra^{0.31}$ for $10^7 < Ra \le 10^9$, quite similar to 3D turbulent data 
that show $Nu\sim 0.105\, Ra^{0.31}$ in that range.  
The optimum solution  
is a simple yet multi-scale coherent solution whose horizontal wavenumber scales as
$0.133 \, Ra^{0.217}$. That solution is unstable to larger scale perturbations 
and in particular to mean shear flows, yet it appears to be relevant as a backbone 
for turbulent solutions, possibly setting the scale, strength and spacing of elemental plumes.

\end{abstract}

\maketitle

Rayleigh-B\'enard convection is the buoyancy-driven motion of a fluid contained between two horizontal plates and heated from below. It is a paradigmatic problem with a rich array of nonlinear physics from deterministic chaos \cite{Lorenz1963} and period doubling \cite{libchaber1982perioddoubling} to pattern formation 
 \cite{bodenschatz2000AnnRev} and turbulence \cite{ahlers2009RMP}.  A fundamental issue is to determine how the heat flux $\HF$ scales with the temperature difference $\Delta T$ between the bottom and top plates. There is a critical  $\Delta T_c$ such that heat transport is by conduction with $\HF \sim \Delta T$ for $\Delta T < \Delta T_c$ but macroscopic fluid motion develops for $\Delta T> \Delta T_c$ and enhances heat transport. Bifurcations take place as $\Delta T$ is increased, leading to turbulent flow and $\HF \sim (\Delta T)^{4/3}$ and perhaps even as high as $\HF \sim (\Delta T)^{3/2}$. Understanding these bifurcations and the actual asymptotic scaling of $\HF$ have been the subjects of many experimental, theoretical and numerical studies focusing on \emph{turbulent} convection. Here, we consider \emph{coherent} convection -- steady flows that may be the permanent form of the coherent structures (\emph{`plumes'}, \emph{`thermals'} and large scale \emph{`winds'}) permeating turbulent convection. 

In Rayleigh-B\'enard convection, the fluid is contained between two infinite horizontal plates and the Boussinesq approximation is made so the governing equations are 
\begin{align}
\pdd{\vel}{t} + ({\vel}\cdot\grad)\vel
 + \grad p &= g \alv T \, \yh+ \nu \nabla^2 \vel, \label{vueqn} \\
\pdd{T}{t} + (\vel \cdot \grad)T &= \kappa \nabla^2 T,
   \label{Teqn1}
\end{align}
for the velocity  $\vel$ and the temperature  $T$. 
Incompressibility \mbox{$\grad\cdot \vel=0$}  is maintained by the kinematic pressure $p$ and $-g\yh$ is the constant acceleration of gravity. The fluid density is $\rho \approx \rho_0 \, (1- \alv T)$ with $\alv  \ge 0$ the volumetric thermal expansion coefficient and $\rho_0$ a constant reference density, $\nu > 0$ is the kinematic viscosity and $\kappa> 0$ the thermal diffusivity.
The bottom plate temperature  is fixed at $\Delta T/2$ and the upper plate is at $ -\Delta T/2$.
The conduction state is the solution $\vel=0$, $T= - y \Delta T /H$
to  \eqref{vueqn} and \eqref{Teqn1}, with $-h \le y \le h= H/2$. 
This is a steady solution for all values of the parameters, but Rayleigh showed that it is linearly unstable when the \emph{Rayleigh} number 
$Ra = g\alv \Delta T\, H^3/{(\nu \kappa)}$
is larger than a critical value $Ra_c$, independent of the \emph{Prandtl} number $Pr = \nu/\kappa$. 
 
 We consider no-slip boundary conditions for which \mbox{$Ra_c \approx 1708$} for horizontal wavenumber\cite{dominguez1984} $\alpha_c=3.116/H$. 
Convection develops and enhances heat transport for $Ra>Ra_c$. For statistically steady solutions, integration of \eqref{Teqn1}  yields the heat flux,\cite{malkus1954B}
\beq \HF =  - \kappa \left.\frac{d\overline{T}}{dy}\right|_{y=\pm h} = \HF_0+ \langle \, v T \, \rangle 
\label{heatflux}
 \eeq
where $\HF_0=\kappa \Delta T/H$ is the heat flux in the absence of fluid motion and $v=\yh\cdot \vel$ is the vertical velocity with $\overline{\,*\,}$ and $\langle * \rangle$ denoting horizontal and domain averages, respectively. 
 
 
 A classic scaling argument is that heat transport is determined by marginal stability of the thermal boundary layers and of the mean temperature gradient in the interior.\cite{malkus1954B}
 A simple version of this argument assumes an isothermal $T=0$ interior with boundary layers of thickness $\delta$ such that
 \footnote{Estimated from linear stability about the conduction state but with free `outflow' boundary conditions at the top edge of the boundary layer, that is $\pd_y v = \pd_y^3 v = \pd_y (T+y)=0$.}
 $Ra^{(\delta)} \equiv  g \alv (\Delta T/2) \delta^3/(\nu \kappa)\approx 1708/16$,
  yielding  heat flux
\beq 
  \HF \sim \kappa \frac{\Delta T/2}{\delta} \approx  \left(\frac{g\alv \kappa^2}{1708\, \nu}\right)^{1/3} (\Delta T)^{4/3} 
\quad \Leftrightarrow \quad 
Nu \sim \frac{h}{\delta} \sim 0.084\, Ra^{1/3}  \label{NuMarginal} 
\eeq 
  independent of $Pr$, where the \emph{Nusselt} number $Nu= {\HF}/\HF_0$.
This scaling also follows from assuming\cite{spiegel1971} that the heat flux becomes independent of the height $H$ as $Ra\to \infty$.

Another simple argument imagines a fluid at temperature $T=0$, with cold plumes at temperature $-\Delta T/2$ free-falling a distance $H=2h$ from top to bottom plates (and hot $\Delta T/2$ plumes `free-rising' from bottom to top) at average speed $V=\sqrt{g'h}$, where $g' = g \alv \Delta T/2$ is the reduced gravity. 
This yields the heat flux
\beq 
 \HF \sim V {\Delta T}/{2} =
\frac{1}{4} \sqrt{g\alv H}\, (\Delta T)^{3/2} 
\quad \Leftrightarrow \quad Nu \sim \frac{Vh}{\kappa} = \frac{1}{4} \left(Ra\, Pr\right)^{1/2}.
\label{NuFreeFall}
\eeq 
This \emph{inertial} scaling also follows from the standard turbulence assumption\cite{spiegel1971} that heat transport becomes independent of $\nu$ and $\kappa$ in the limit
$Vh/\nu\to \infty$, $Vh/\kappa \to \infty$. 

Kraichnan's mixing length theory\cite{Kraichnan1962} yields various scalings in distinct regions of the $(Ra,Pr)$ parameter space. He obtains
$Nu \sim Ra^{1/3}$, for fixed $Pr$, 
 but predicts a transition to $Nu \sim  Ra^{1/2}(\ln Ra)^{-3/2}$
for very large $Ra \gtrsim 10^{12}$ when the shear boundary layers would be turbulent. 
Grossmann and Lohse's comprehensive scaling theory\cite{grossmann2000scaling} incorporates classic Kolmogorov scaling of velocity and temperature dissipation rates in the bulk together with thermal and viscous boundary layers. Their theory 
fits many experimental data sets.\cite{stevens2013unifying}  
  They predict a transition to an `ultimate' regime  similar to Kraichnan's for $Ra \gtrsim 10^{14}$ and argue that their log correction would yield an effective scaling\cite{GrossmannLohse2011} of about $Nu\sim Ra^{0.38}$ in the range $10^{12} < Ra < 10^{15}$.  Some  experimental results show transition to an ultimate regime\cite{Chavanne1997,Chavanne2001, Roche2010, He2012}   while others do not.\cite{Niemela2000,NiemelaSreeni2006}
 However, all the data is well-fitted by $Nu \sim 0.105 \, Ra^{0.31}$ for $Ra< 10^{11}$ and that is the regime for which we report on unstable coherent states, in particular optimum transport solutions with  $Nu \sim 0.115\, Ra^{0.31}$.

Rigorous upper bounds on heat transport for no-slip boundary conditions \cite{Howard1963,Busse1969,DoeringConstantin1996, Kerswell2001, Plasting2003} yield 
\mbox{$Nu-1 \lesssim 0.026\, Ra^{1/2}$} as $Ra \to \infty$, for any $Pr$, showing that the free-fall scaling \eqref{NuFreeFall} cannot hold for large $Pr$.  Rigorous bounds for free-slip\cite{WhiteheadDoering2011,WCKD2015} yield $Nu-1 \lesssim 0.106 \,Ra^{5/12}$   that conflicts with \eqref{NuFreeFall}  for any $Pr$.

We consider 2D flow with $\vel= u(x,y,t) \xh + v(x,y,t) \yh$ and use $h=H/2$, $V=\sqrt{g'h}$, $\tau=h/V$, $\Delta T/2$ as our characteristic length, velocity, time and temperature scales, respectively.
Eliminating $p$ by taking the $\yh$ component of the curl of the curl of \eqref{vueqn}   yields
 \begin{align}
\pd_t \nabla^2 v &= {\nu} \; \nabla^2 \nabla^2 v +  \pd_x^2 T + \pd_x  \left(v \nabla^2 u - u \nabla^2 v \right), \label{v2deqn}\\[1ex]
\pd_t T&=  {\kappa} \; \nabla^2 T -  (\vel \cdot \nabla)T , \label{Teqn2} \\[1ex]
 \pd_t \ubar & = {\nu} \; \pd_y^2 \, \ubar - \pd_y \overline{uv}  \label{ubareqn}
\end{align}
where $\nu$ and $\kappa$ are now non-dimensionalized by $Vh=\sqrt{g'h^3}$ and  relate to the  Rayleigh and Prandtl numbers as 
    \beq {\nu} =4 \sqrt{\frac{Pr}{Ra}}, \quad  \kappa=\frac{4}{\sqrt{Ra \,Pr}}.  \label{nuokapo}\eeq
All results in this paper are for $Pr=7$ (water). 
The horizontal velocity $u$ is obtained from $v$ and $\pd_x u + \pd_y v=0$ except for its horizontal average $\ubar(y,t)$ that is determined by \eqref{ubareqn}, where the overbar denotes a horizontal average. Equations \eqref{v2deqn}, \eqref{Teqn2}, \eqref{ubareqn} are considered with no-slip boundary conditions 
  \beq v=0, \; \pd_y v = 0, \; T=\mp 1 \quad \text{at} \quad y=\pm 1, \eeq
together with periodicity of period $L=2\pi/\alpha$ in the $x$ direction.  

\begin{figure}[t]
\includegraphics[width=8.6cm]{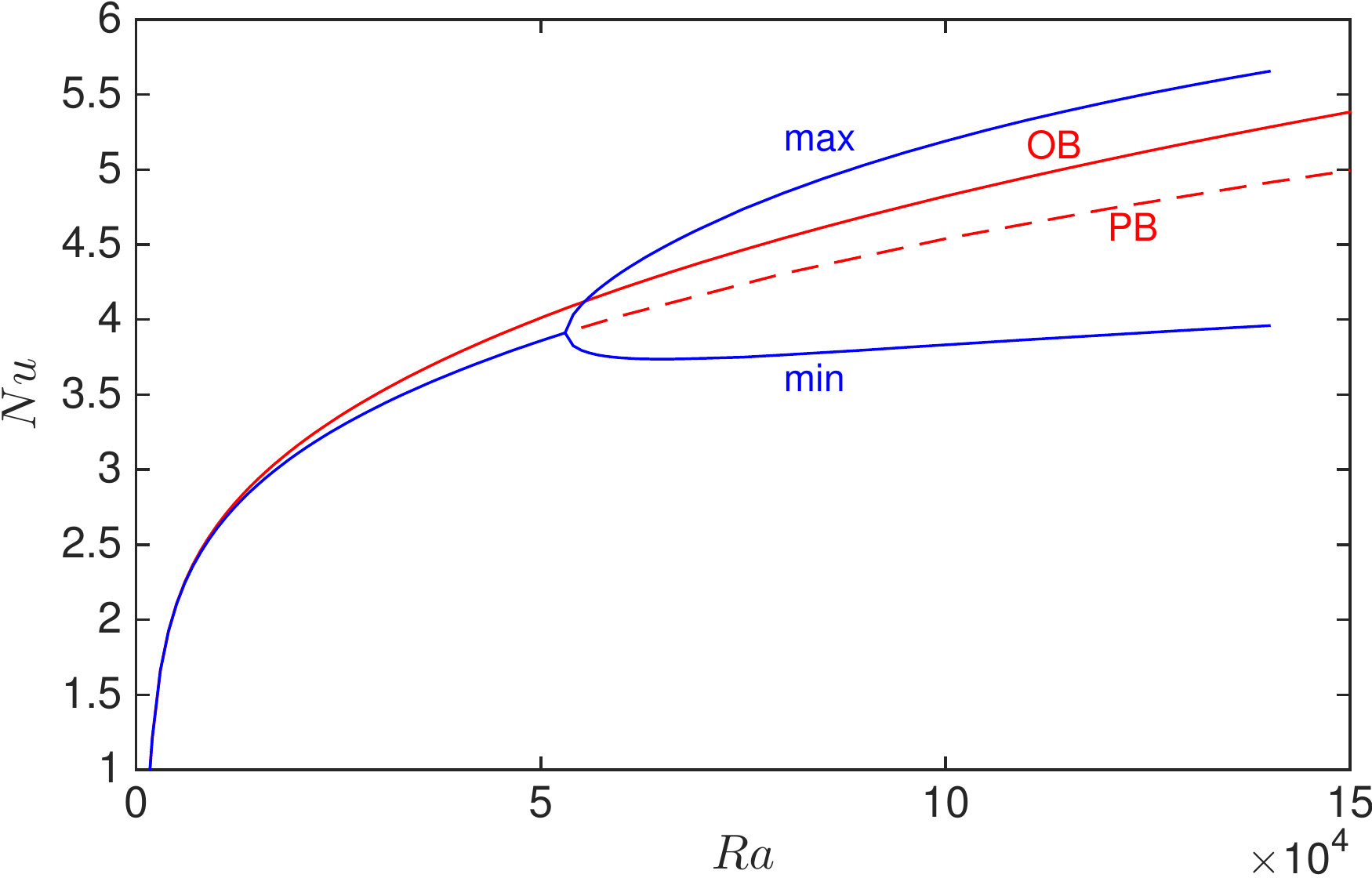}
\caption{$Nu$ vs.\ $Ra$ for steady primary branch (PB) with $L/H=2$ and optimum branch (OB) bifurcating from conduction state at $Ra=1708$, $Nu=1$. Primary branch bifurcates to a time-periodic solution near $Ra=53\, 000$, max and min $Nu$ achieved by the periodic solution are plotted. Unstable steady state is dashed.}
\label{RaNuLIN}
\end{figure}

\begin{figure}[t]
\includegraphics[width=8.6cm]{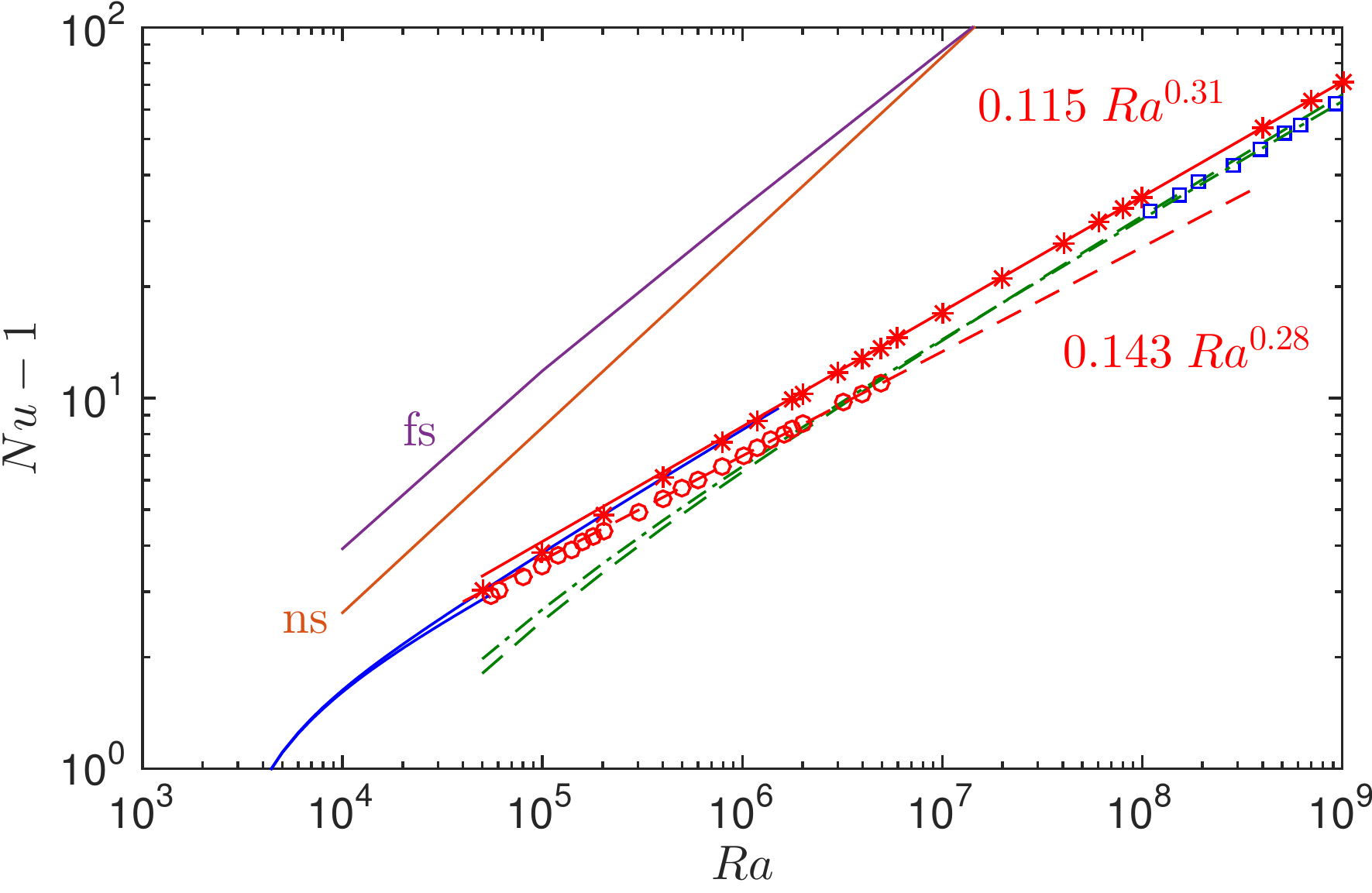}  
\caption{$Nu-1$ vs.\ $Ra$ for \emph{primary branch} ($\alpha=\pi/2$, red $\circ$'s) with $Nu-1 \sim 0.143\, Ra^{0.28}$ (red dash) and \emph{optimum branch} (red $\ast$'s) with $Nu-1 \sim 0.115\, Ra^{0.31}$ (least square fit in $10^7 <  Ra \le  10^9$, red solid). 
Lower (green) dash is the 3D turbulent data fit\cite{NiemelaSreeni2006}  $Nu \sim  0.088\, Ra^{0.32}$ , 
(green) dash-dot is the 3D turbulent data fit\cite{He2012} $Nu \sim 0.105 \,Ra^{0.312}$, both for domain aspect ratio 1/2. The blue $\square$'s  for $Ra>10^8$ is the aspect ratio 4 data\cite[Table 1, $Nu_{corr}$]{NiemelaSreeni2006}. 
Line `fs' is the best free-slip upper bound\cite{WCKD2015} $Nu-1\lesssim 0.106\, Ra^{5/12}$.
Line `ns' is the best no-slip upper bound\cite{Plasting2003} $Nu-1 \lesssim 0.02634\, Ra^{1/2}$. }
\label{RaNuLOG}
\end{figure}

\begin{figure}[t]
\includegraphics[width=8.6cm]{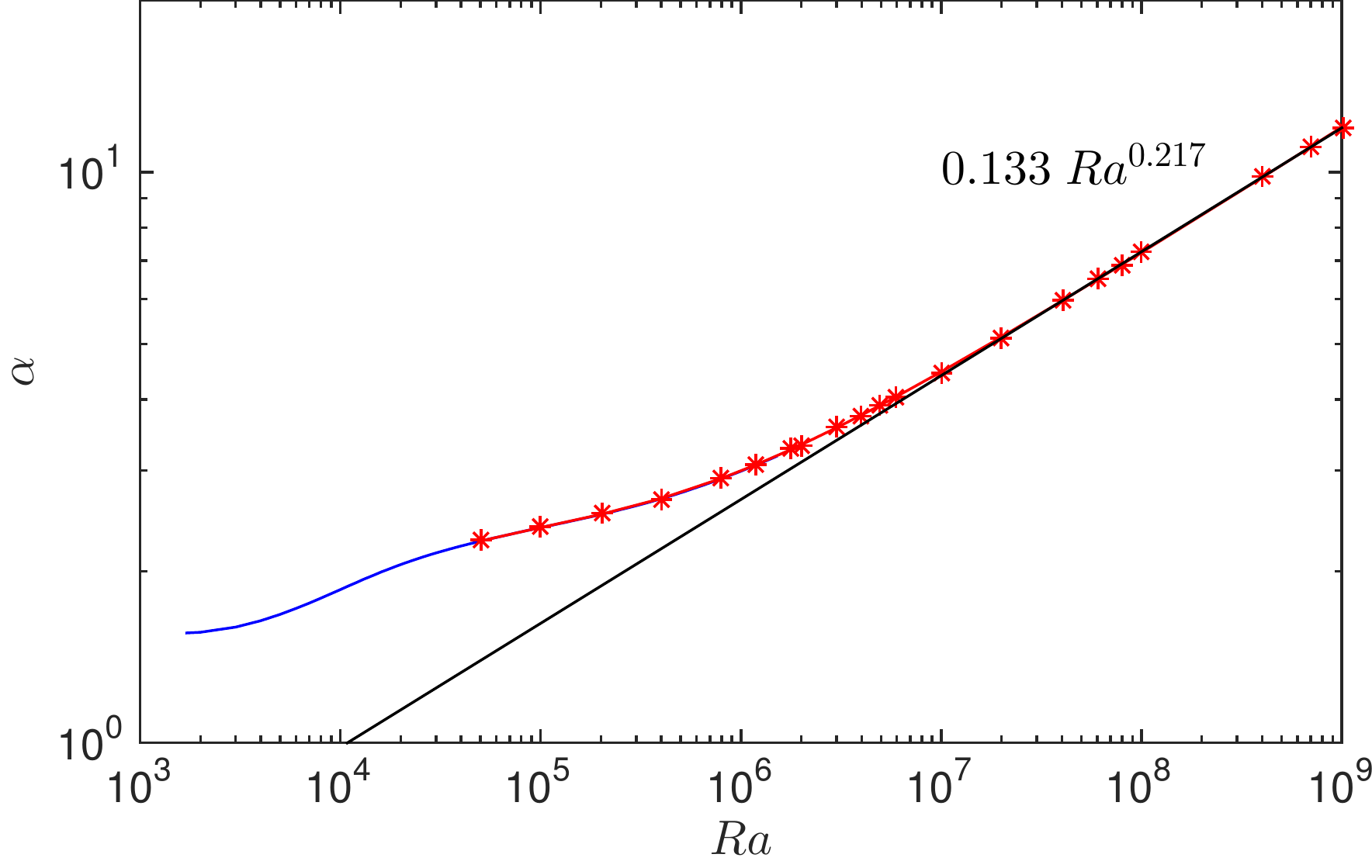}  
\caption{Horizontal wavenumber $\alpha=\alpha^{opt}(Ra)$ that maximizes heat flux $Nu$, $\alpha^{opt} \approx 0.133 \, Ra^{0.217} $(least square fit in $10^7 <  Ra \le  10^9$).}
\label{alphaRaLOG}
\end{figure}

We look for \emph{steady} solutions, $\pd_t=0$, that bifurcate from the conduction state at $Ra \approx 1708$ for wavenumber $\alpha \approx 3.116/2$ corresponding to aspect ratio $L/H  \approx 2.016 $. 
Those convective solutions obey mirror symmetry
\beq [u,v,T](x,y) =  [-u,v,T](-x,y) \label{mirror}
\eeq
as well as shift-reflect symmetry
\beq [u,v,T](x,y) = [u,-v,-T](x+\frac{L}{2},-y). \label{shiftreflect}
\eeq 
Equations \eqref{v2deqn}, \eqref{Teqn2} are discretized using a Fourier expansion in $x$ and Chebyshev integration \cite{W03}  in $y$ with time-marching to steady states. Two distinct codes have been written, code 1 uses a 3rd order time-accurate scheme,\cite{SpalartMoserRogers1991} code 2 uses a semi-implicit forward-backward Euler time discretization together with Chebyshev tau of the 2nd kind \cite{CharalambidesWaleffe2008} for the $v$ equation. Code 2 typically uses inconsistent time integration, with smaller time steps for smaller wavenumbers and for the $v$ equation than for the $T$ equation, to speed up or enable convergence to steady state.  Both codes are dealiased in $x$ and $y$ using the 2/3 rule. The results of both codes overlap or connect very well as seen in figs.\ \ref{RaNuLIN}, \ref{RaNuLOG}, \ref{alphaRaLOG}, discussed below. The results also match a 3rd code based on finite differences and a Newton-Krylov iteration to steady state\cite{SondakSmithWaleffe2015} (not shown here). Mirror symmetry \eqref{mirror} is imposed and eliminates the mean flow \eqref{ubareqn}. Shift-reflect symmetry \eqref{shiftreflect} is not imposed explicitly but is satisfied by the steady solutions presented here.

We show two branches of nonlinear steady solutions that bifurcate from the conduction state at $Ra\approx 1708$, $\alpha\approx 1.558$. The \emph{primary branch} has fixed horizontal wavenumber $\alpha$  (rounded here to $\alpha=\pi/2$ $\Leftrightarrow L/H=2$). The \emph{optimum branch} adjusts  $\alpha$ to maximize heat flux $Nu$. The $(Ra,Nu)$ curves for both solutions are shown in figures \ref{RaNuLIN} and  \ref{RaNuLOG}. The steady primary solution is stable up to $Ra\approx 53\,000$ where it spawns a time-periodic solution. The maximum and minimum $Nu$ achieved by that periodic solution are shown in fig.\ \ref{RaNuLIN}, that time periodic solution was computed with the time-accurate code 1. The unstable steady state was continued with code 2 up to $Ra=5.\, 10^6$ (symbols $\circ$ in fig.\ \ref{RaNuLOG}).
Mirror symmetry \eqref{mirror} suffices to stabilize the optimum solution in its fundamental periodic domain. That optimum solution was calculated up to $Ra=1.4\, 10^6$ with code 1 and $Ra=10^9$ with code 2 (symbols $\ast$ in fig.\ \ref{RaNuLOG}).  The optimum solution at $Ra=10^9$ is well-resolved with 200 Chebyshev polynomials in $y$ and 200 Fourier modes in $x$, after dealiasing.
 

\begin{figure}[t]  
\centering
\includegraphics[width=8.6cm]{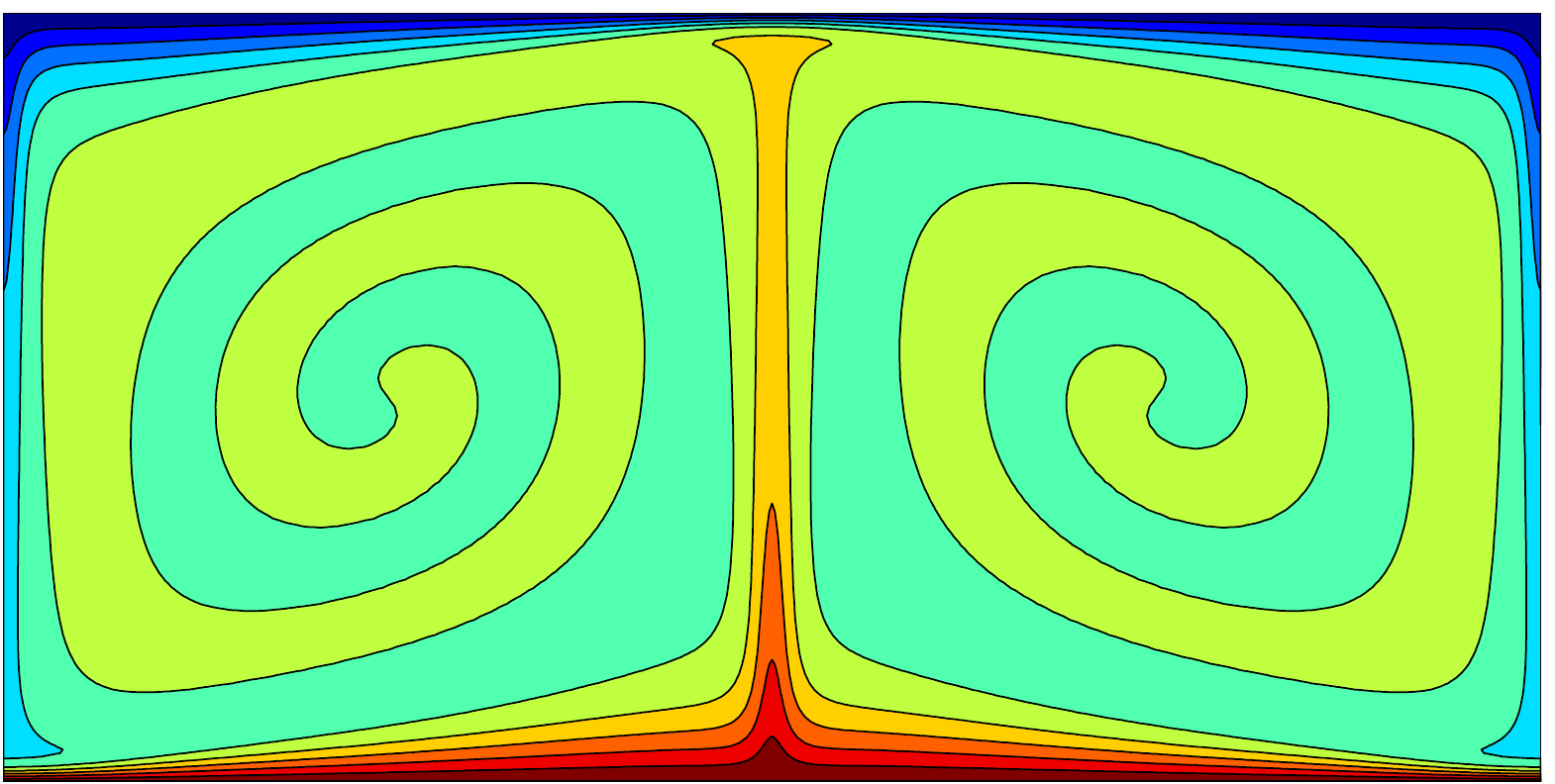}\\
\includegraphics[width=8.6cm]{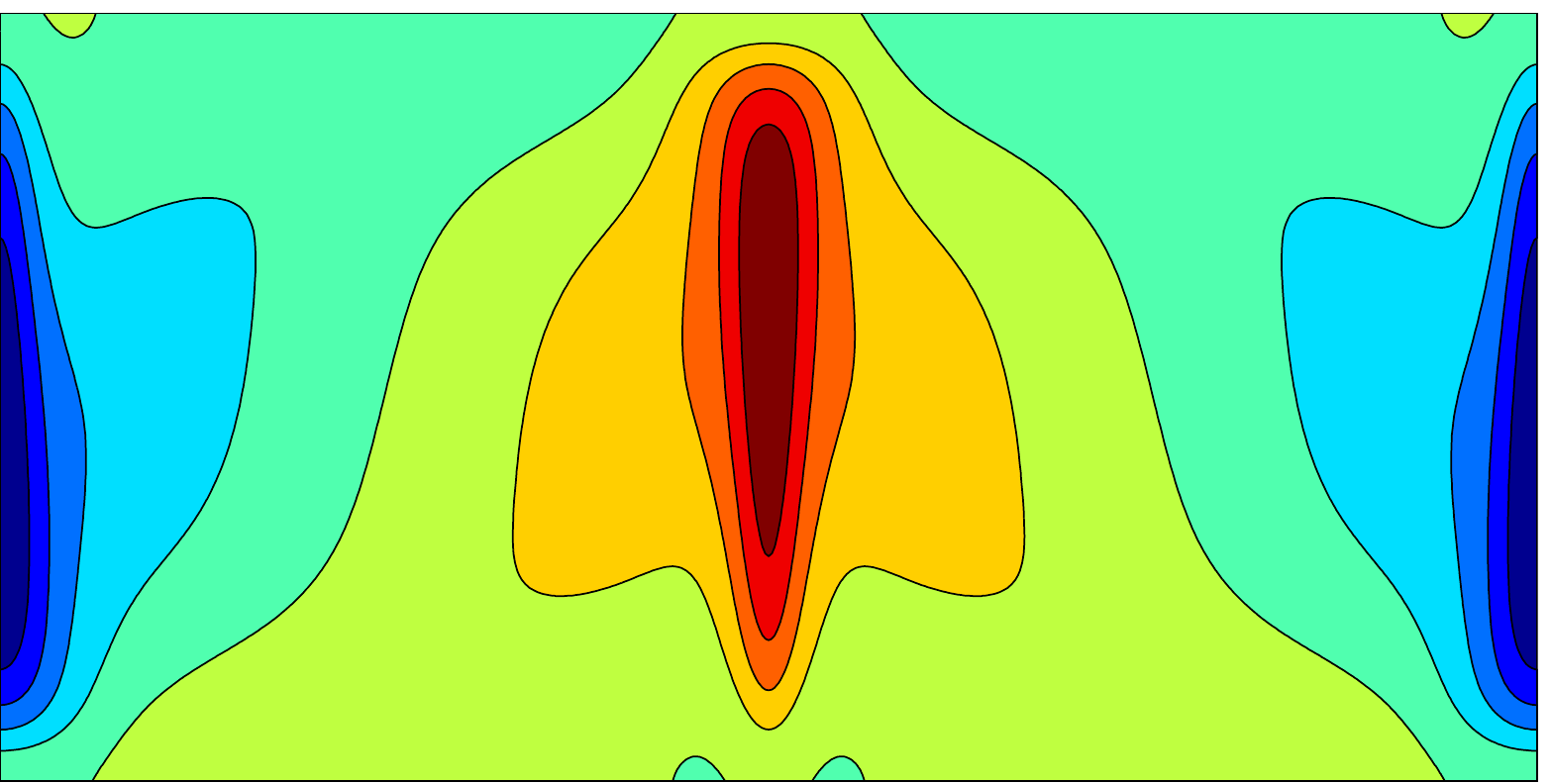}\\
\caption{\emph{Primary solution.} Temperature $T$ (top) and vertical velocity $v$ (bottom) at $Ra=5.\ 10^6$ for $L/H=2$, $Nu=11.93$. Equispaced contours at 10\% of max-min, actual aspect ratio with $-2 \le x \le 2$ horizontal and $-1 \le y \le 1$ vertical.}
\label{Ra5MPrimary}
\end{figure}

 The heat flux for the unstable primary solution scales as $Nu-1 \approx 0.143\, Ra^{0.28}$ according to a least square fit in $5. 10^5 \le Ra \le 5. 10^6$. That solution develops a delicate spiral structure in the temperature field but not in the velocity (fig.\ \ref{Ra5MPrimary}). The winding of these temperature spirals continuously increases with Rayleigh number. 
 Convergence of our algorithm to that unstable steady solution becomes quite slow for increasing $Ra$, apparently because of the center region of these spiral structures where $v, T \approx 0$.

 The optimum solution increases wavenumber $\alpha$ with $Ra$ as $\alpha \approx 0.133 \, Ra^{0.217}$ (fig.\ \ref{alphaRaLOG})
 to achieve an optimum heat flux $Nu-1 \approx 0.115 \, Ra^{0.31}$  (fig.\ \ref{RaNuLOG}),  from a least square fit in $10^7 < Ra \le 10^9$. This optimum heat flux scaling is quite similar to the heat flux observed in 3D turbulent convection experiments. Indeed, the optimum branch in fig.\ \ref{RaNuLOG} is only slightly above 
 the cryogenic helium gas data fit\cite{NiemelaSreeni2006} $Nu\approx 0.088 \, Ra^{0.32}$   and the pressurized $SF_6$ gas data fit\cite{He2012} $Nu \approx 0.105\, Ra^{0.312}$  (both in a cylinder of aspect ratio 1/2 and both for $Ra<10^{11}$).
 The aspect ratio 4 data\cite[Table 1]{NiemelaSreeni2006} is well-fitted by $Nu-1 \approx 0.102\, Ra^{0.31}$ in $10^8 \le Ra \le 10^{10}$ and lies just below the optimum transport data with $Nu-1 \approx 0.115\, Ra^{0.31}$. This close agreement is remarkable given the differences in dimension, 2D optimum vs.\ 3D data, and Prandtl number, $Pr=7$ for the 2D optimum vs.\ $Pr\approx 0.7$ in the helium gas experiments.
 
\begin{figure}[t] 
\centering
\includegraphics[height=44truemm]{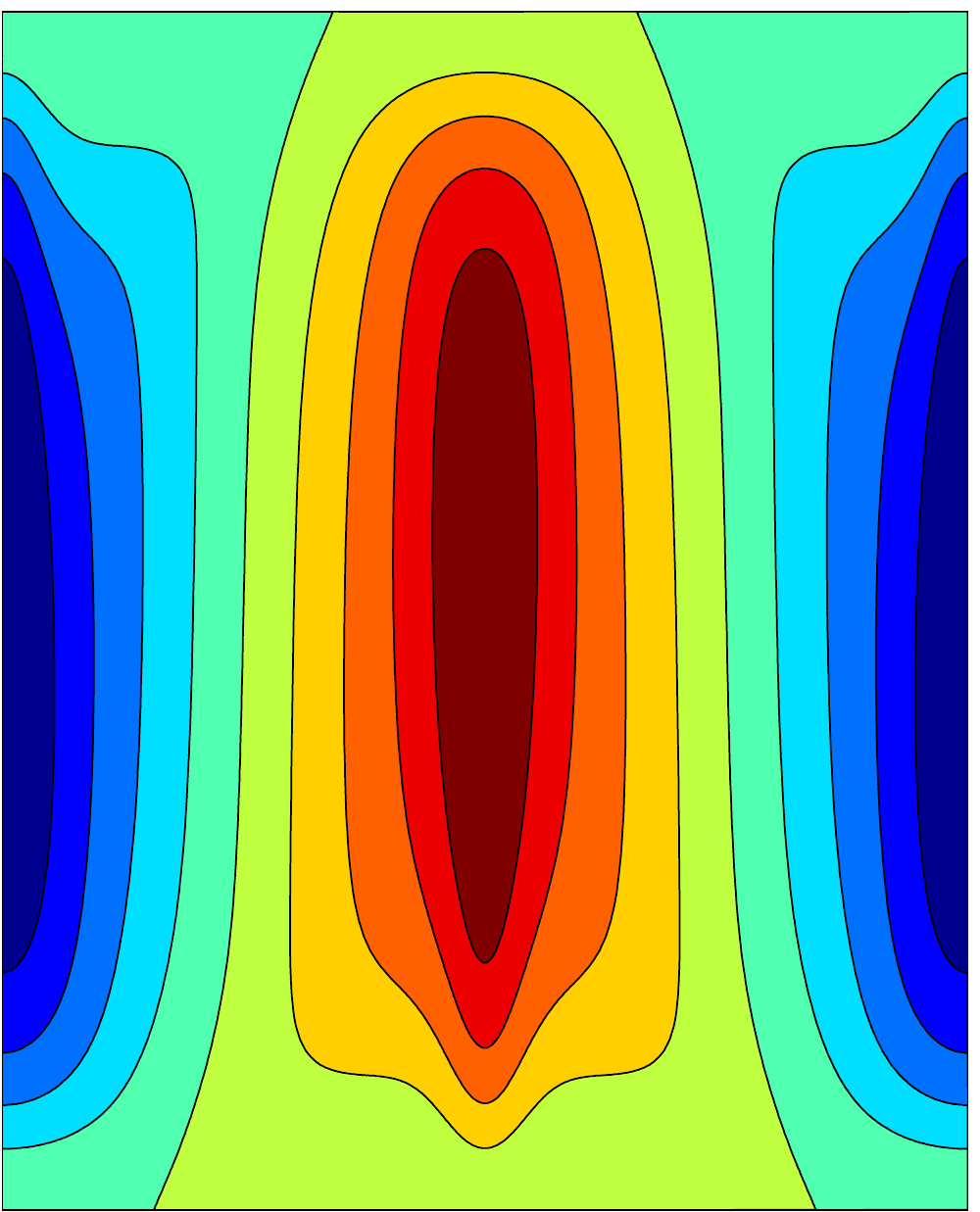}
\includegraphics[height=44truemm]{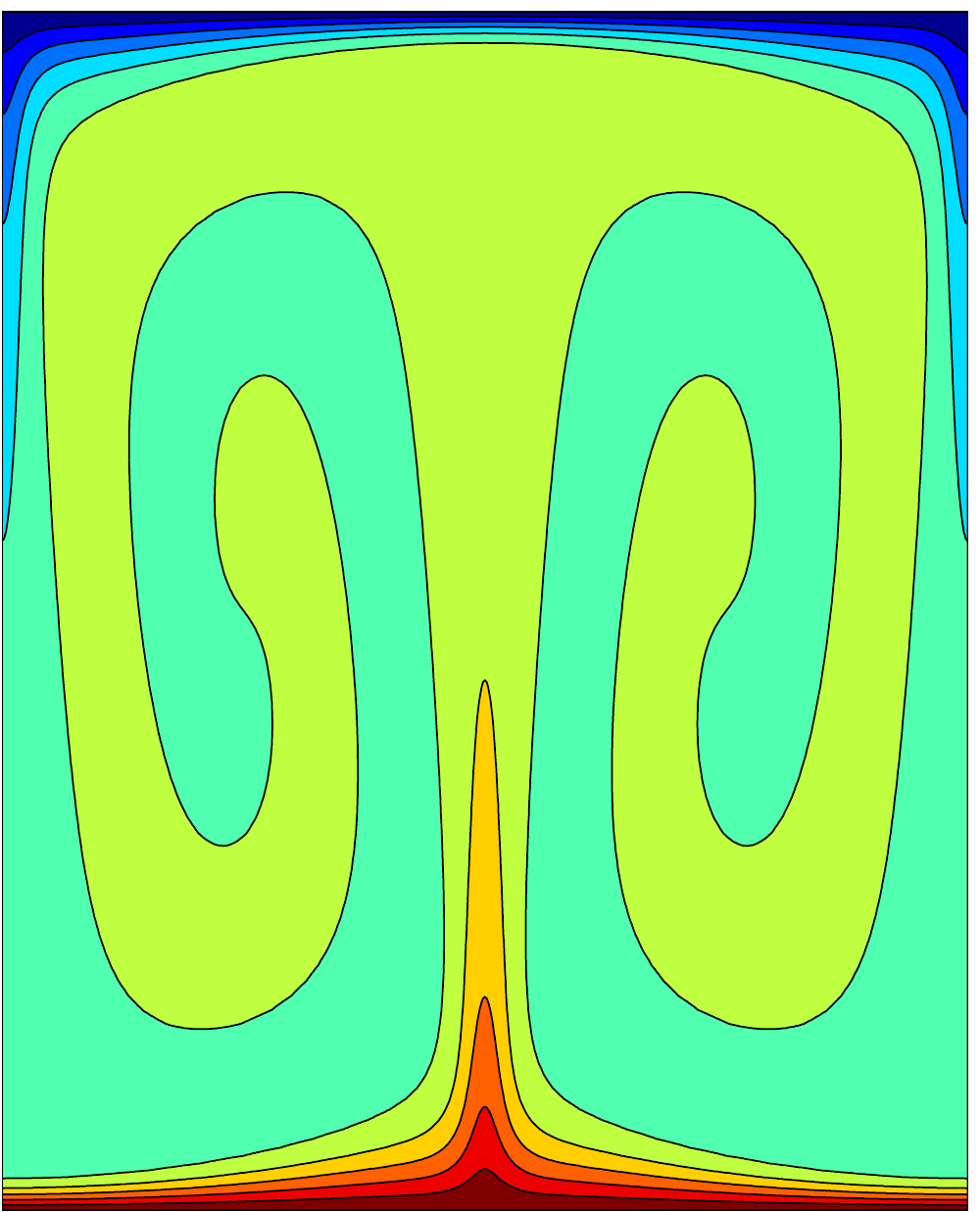}
\includegraphics[height=44truemm]{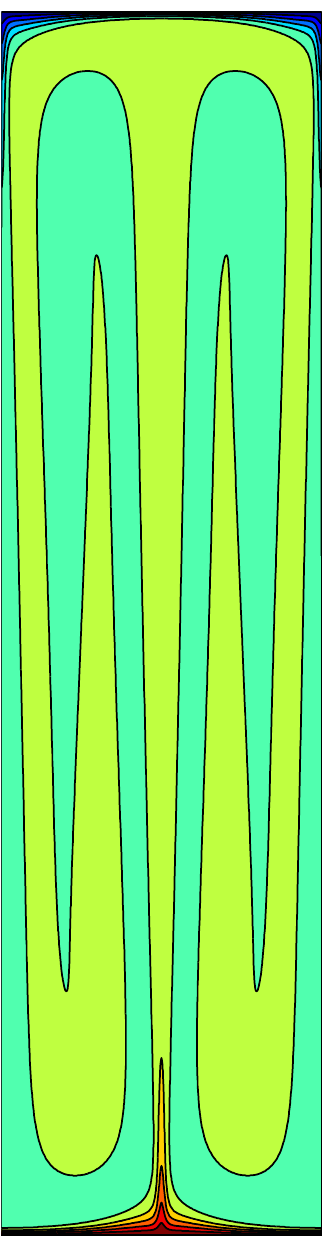}
\caption{\emph{Optimum solution.} Velocity $v$ (left) and temperature $T$ (center) at $Ra=5\, 10^6$, $L/H=0.803$, $Nu=14.72$.
Temperature $T$ (right) at $Ra=10^9$, $L/H=0.262$, $Nu=72.3$.  Equispaced contours at 10\% of max-min, actual aspect ratio.} 
\label{Ra5MOpti}
\end{figure}

The optimizing wavenumber $\alpha=\alpha^{opt}(Ra)$ (fig.\ \ref{alphaRaLOG}) shows an undulation between $1708 < Ra \lesssim 10^6$ that is linked to a temperature spiral structure in the optimum transport solution as well (fig.\  \ref{Ra5MOpti}). A temperature updraft of hot fluid (and downdraft of cold fluid) develops between $Ra=1708$ and $Ra \approx 10^4$, but a spiral structure with slight downdraft of warm fluid (and updraft of cool) develops in  $10^4 \lesssim Ra \lesssim10^5$,  corresponding to the bump in the curve $\alpha=\alpha^{opt}(Ra)$ in fig.\ \ref{alphaRaLOG}. The warm spiral begins winding back up at $Ra \approx 10^5$. The spiral does not appear to ever wind back down for higher $Ra$, developing instead an increasingly jagged structure (fig.\ \ref{Ra5MOpti}, right), and $\alpha^{opt}$  approaches the power law scaling $\alpha^{opt}\simeq 0.133\, Ra^{0.217}$. 
The structure of optimum $Nu$ solutions depends on Prandtl number $Pr$. This is investigated in forthcoming work\cite{SondakSmithWaleffe2015} which also shows that the locally optimum steady solution discussed here is in fact the global optimum over $\alpha$.

The close agreement between the optimum transport 2D steady state solutions and 3D turbulent data in fig.\ \ref{RaNuLOG} is intriguing. Although this agreement could be fortuitous, it strongly suggests that a single unstable steady solution may capture key statistical features of fully developed turbulent flows, such as the net heat flux and the mean temperature profile as well as the strength and scale of elemental plumes,  as in shear flows \cite{W03,kawahara_AnnRev2012}. A search for such maximum momentum transport solutions in shear flows was initiated over 10 years ago but not completed because of the higher computational complexity of those 3D solutions (J.\ Wang and F.\ Waleffe, 2004, unpublished). 

Our results are connected with Malkus' theory of turbulent convection \cite{malkus1954B} and subsequent work on upper bounds \cite{Howard1963,Busse1969, WhiteheadDoering2011,WCKD2015,DoeringConstantin1996,Kerswell2001,Plasting2003}. Two key ingredients of Malkus' theory are maximum heat transport and marginal stability of the mean temperature profile and the smallest scales of motion. Both ingredients are included in our calculations that maximize heat transport over horizontal wavenumber $\alpha$ and track steady state solutions that bifurcate from the marginal stability critical point at $Ra\approx 1708$, $\alpha\approx 1.558$. We conjecture that our 2D results are in fact the 3D optimum  transport solutions (for infinite or periodic horizontal directions). The upper bound results also assume 2D optimizers.  Whether the ultimate scaling of these optimum Boussinesq solutions is $Nu \sim Ra^{1/3}$  as $Ra \to \infty$ remains to be seen but is possible, as is an abrupt transition to a smaller scale optimum solution.\cite{SondakSmithWaleffe2015}

Why the optimum transport solution should capture gross 3D turbulence characteristics might be understood as a `winner-take-all' effect, 
where the optimum solution consumes all available potential energy so no other flow can be sustained. 
The optimum solution appears to be stable when mirror symmetry \eqref{mirror} is imposed and length scales are restricted to be less or equal to the optimum wavelength. The optimum solution is unstable to larger scale perturbations, in particular to subharmonics where plumes merge and form bigger plumes in a cyclic or quasi-cyclic fashion. It is also unstable to a mean shear flow \eqref{ubareqn} when mirror symmetry is allowed to be broken. These instabilities would be the source of the `turbulence' but the underlying unstable coherent solutions control the heat transport.

\begin{acknowledgments}
The authors would like to thank Charles Doering (U Michigan) and David Sondak (UW Madison) for helpful comments and discussions and Detlef Lohse (U Twente) for providing many recent references on the ultimate regime and the Grossmann-Lohse scaling theory. This research was partially supported by NSF grants DMS-0807349 (FW, AB) and DMS-1008396 (AB, LMS).
\end{acknowledgments}


\bibliography{\biblio}

\end{document}